\documentclass[12pt]{article}
\usepackage{epsfig,latexsym,colordvi}
\usepackage{graphicx}

\begin{document}
\bibliographystyle{h-physrev4}

\begin{center}
{\LARGE Topological Fluctuations in Dense Matter \\with Two Colors}

\vskip 0.3cm
{\bf Simon Hands and Philip Kenny}

\vskip 0.3cm
{\em Department of Physics, College of Science, Swansea University,\\
 Singleton Park, Swansea SA2 8PP, UK}
\end{center}

\noindent
{\bf Abstract:}
We study the topological charge fluctuations of an SU(2) lattice gauge theory
containing both $N_f=2$ and 4 flavors of Wilson fermion, at low temperature
with non-zero chemical potential $\mu$. The topological susceptibility,
$\chi_T$, is used to characterize differing physical regimes as $\mu$ is varied
between the onset of matter at $\mu_o$ and and color deconfinement at $\mu_d$.
Suppression of instantons by matter via Debye screening is also investigated,
revealing effects not captured by perturbative predictions.  In particular, the
breaking of scale invariance leads to the mean instanton size $\bar\rho$ becoming
$\mu$-dependent in the regime between onset and deconfinement, with a scaling 
$\bar\rho\propto\mu^{-2}$ over the range $\mu_o<\mu<\mu_d$, resulting in an
enhancement of $\chi_T$ immediately above onset.

\section{Introduction}
Lattice studies of matter at non-zero baryon density are hampered by the `sign
problem', which arises when a quark chemical potential term $\mu$ is included in
the Euclidean QCD action. The resulting complex nature of the fermion
determinant precludes a positive definite probability measure and computational
techniques based on importance sampling break down. A gauge theory which is
accessible to Monte Carlo simulations is QC$_2$D, based on gauge group SU(2),
describing ``two color matter''. In QC$_2$D, quarks belong in the pseudoreal
\textbf{2} representation of SU(2) which can guarantee a positive definite
measure.

Studies of two color matter have been performed utilising a number of fermion
formulations. The series of works obtained from simulations involving two and
four flavors of Wilson fermion~\cite{Hands:2006ve,Hands:2010gd,Hands:2011ye}
have revealed a scenario in which, as $\mu$ is increased, baryonic matter forms
at an onset $\mu_o=m_\pi/2$ whereupon the matter then exists in a superfluid
state with a progression from a dilute gas of tightly-bound diquark pairs to
degenerate quark matter, culminating in color deconfinement at around
$\mu\approx1.1m_\pi$. This Letter supplements this picture with an
investigation of topological effects observed on the same lattice
configurations.

The topological charge density $q_T$ may be defined in terms of the Yang-Mills field tensor as
\begin{equation}
q_T=\frac{1}{32\pi^2}F_{\mu\nu}\tilde{F}_{\mu\nu}\label{eqn:qtop} 
\end{equation}
with $\tilde{F}_{\mu\nu}=\frac{1}{2}\varepsilon_{\mu\nu\rho\sigma}F_{\rho\sigma}$. The
action is minimised when the condition $F_{\mu\nu}=\pm\tilde{F}_{\mu\nu}$ is
satisfied.  The observable measured to study topological charge fluctuations is
the \emph{topological susceptibility}, $\chi_T$, 
defined as 
\begin{equation}
\chi_T=\frac{\left<Q^2\right>}{V},  \label{eqn:chitop} 
\end{equation} 
where $Q=\int d^4x q_T$ and $V=\int d^4x$. 
Using large-$N_c$ methods $\chi_T$ is
estimated by means of the Witten-Veneziano
formula~\cite{Witten:1979vv,Veneziano:1979ec} 
\begin{equation}
\chi_T={f_\pi^2\over{2N_f}}\Bigl(m_\pi^2+m_{\eta^\prime}^2-2m_K^2\Bigr)
\label{eq:WV}
\end{equation}
to be (180 MeV)$^4$ in the SU(3)
gauge vacuum. Simulations of hot two color matter with two flavors of staggered
quark (equivalent to $N_f=8$ continuum quark flavors) have shown this quantity
drops sharply at the deconfining temperature and have suggested this also
happens at non-zero chemical potential
~\cite{Alles:2006ua,Alles:2006ea}. When $\chi_T$
is measured as a function of $a\mu$, the susceptibility remains constant before
dropping dramatically at a critical chemical potential corresponding to both
deconfinement and chiral symmetry restoration.

In a semi-classical picture toplological charge is localised on four-dimensional
objects called instantons, which are solutions of the sef-dual condition for 
a local minimum of the action~\cite{Belavin:1975fg}. 
Another observable of interest is the size of an instanton $\rho$. This is a
measure of the extent to which the gauge field action is localised.
For classical Yang-Mills instantons
the size may be considered arbitrary due to scale invariance and so $\rho$ does
not depend upon the action, and vice versa. However, in the quantum vacuum scale
invariance is broken, and the typical size of an
instanton is estimated to be in the region of 0.3
fm~\cite{Shifman:1978zp,Shuryak:1981fza}.

In dense matter, Debye screening of color charge leads to instanton
suppression~\cite{Schafer:1998up}. Perturbative
calculations~\cite{Shuryak:1982hk} predict that instanton number at large
chemical potential should go like 
\begin{equation}
n(\mu)=n(\mu=0)\exp\left(-N_f(\rho\mu)^2\right).  \label{eqn:flavsup}
\end{equation} 
Therefore, as the number of quark flavors $N_f$ is increased, 
instantons should be suppressed and $\chi_T$ should decrease. It
should also be expected that, if the average instanton size $\rho$ is indeed 
fixed, then the extra matter present as $\mu$ is increased will
screen the topological charge and suppress $\chi_T$ still further.

\section{Methodology}
In order to explore instanton effects on a lattice we replace the continuum
topological charge density $q_T$ (~\ref{eqn:qtop}) with its lattice
counterpart 
\begin{equation}
q_L(x)=\frac{1}{32\pi^2}\epsilon_{\mu\nu\rho\sigma}\mbox{Tr}\left(U_{\mu\nu}(x)U_{\rho\sigma}(x)\right)
\label{eqn:qldefine} \end{equation} 
where $U_{\mu\nu}(x)$ is the product of link
variables around a plaquette at site $x$ in the $\mu-\nu$
plane~\cite{DiVecchia:1981qi}. The charge density is thus measured by taking the
trace of the product of two orthogonal plaquettes. The total charge $Q_L$
is obtained via $Q_L=\sum_x q_L(x)$.  Within each configuration, the
peaks due to the presence of instantons (whose structure may  extend over a
scale $\rho\gg a$, where $a$ is the lattice spacing) 
are mutated by short scale ($\mathcal{O}(a)$) fluctuations.
Such UV fluctuations are highly undesirable as they contribute to the total
charge but obscure the `real' instantons, and so the measured susceptibility
can be an overestimate~\cite{Smith:1998wt}. The lattice topological susceptibility
$\chi_L\equiv\langle Q_L^2\rangle/V$ 
differs from the continuum value by both a multiplicative factor $Z$
and an additive one $M$: 
\begin{equation}
\chi_L=Z^2a^4\chi_T+M.
\end{equation}
$Z$ and $M$ depend on several factors including the quark mass, the inverse
coupling $\beta$ and the choice of fermion operator~\cite{Alles:2006ea}. In
general, on the lattice, $Z\neq1$ and the charge $Q_L$ is not integer-valued. The
challenge is to minimise the unwanted, short distance contributions while
in the process
recovering the continuum value in an unambiguous fashion. 

$Q_L$ for a given configuration of gauge fields is calculated by means of
Eqn.~(\ref{eqn:qldefine}). The effects of UV fluctuations are minimised by
cooling~\cite{Teper:1985rb}, whereby a new configuration is generated from the old
by visiting lattice sites in turn and minimising the action locally. 
Repeating this succesively
has the effect of smoothing out fluctuations and revealing the
underlying topological structure in the gauge fields. 
By prudent use of cooling, the multiplicative factor
$Z\rightarrow1$ as the unwanted fluctuations are eliminated. However, excessive cooling
eliminates not just the UV fluctuations but will also shrink and
ultimately eradicate the `real' instantons. If cooling shrinks an instanton
until its size $\rho<a$ then it `falls through' the lattice and some of the
topological information is lost. If only larger instantons contribute to
the total charge then there is a tendency to underestimate $Q_T$. Information can
also be lost as too much cooling has a tendency to annihilate instanton -- 
anti-instanton pairs. The total charge may remain the same but the charge
density is reduced. Therefore, it is vital that good control of the cooling
process is maintained.

The additive constant $M$ may be dealt with by equating it to the value of the topological susceptibility in the $Q_T=0$ sector, setting $M=\chi_0\equiv\chi_T(Q=0)$. As we have no prior knowledge to suggest that our ensemble is in the trivial sector we must modify Eqn. (\ref{eqn:chitop}). In the non-trivial sector $M$ can be eradicated by redefining
\begin{equation}
a^4\chi_T=\frac{\left<Q^2\right>-\left<Q\right>^2}{V}
\label{eqn:topsus}
\end{equation}
Thus, by measuring the charges on a number of cooled field configurations with $Z\sim1$ and calculating $\chi_T$ by means of (\ref{eqn:topsus}), the physical topological susceptibility can be extracted from the lattice one. Henceforth, we  discard the references to lattice values via our subscripts $L$ and merely label $\chi$ and $Q$ with the subscript $T$.

The cooling method employed here uses a computer program to read the gauge
field information from each configuration and then calculate the total action
by summing over the plaquettes. In general, this is not the minimum action. A
point is then chosen and a link variable $U_\mu(x)$ is selected. There are 6
plaquettes with this link in common. The code sums the link products, in the
form of unitary matrices which form the `staples' bordering the link $U_\mu(x)$,
resulting in a $2\times2$ matrix $V$. The matrix $V$ is non-unitary and must be
renormalised as $\tilde{V}=(\mbox{Det }V)^{-1/2}V$. Keeping 
$\tilde{V}$ fixed, the action is then minimised by modifying $U_\mu(x)$. 
By systematically working through the old
configuration and updating all links $U_\mu(x)$ a new configuration is produced with a
lower action than the original one. 
This completes the first cooling sweep. By predetermining the number
of sweeps to be performed, the process repeats automatically and the
configuration is cooled to the required extent. When cooling is complete, the
code then searches through the final configuration to find where the peaks of
the action are located and $F\tilde{F}$ at these points is recorded. Setting a minimum cutoff for $F\tilde{F}$ allows the code to disregard the smallest fluctuations. Imposing a second cutoff for the maximum extent of the gauge fields inside an instanton minimizes any finite volume effects associated with excessively large instantons. Once the required topological information is extracted from the
cooled configuration, the program then moves onto the next configuration in the
ensemble and repeats as necessary.

To find the total topological charge on each configuration, a second
program obtains the net value of all the peaks of $F\tilde{F}$ from the output
of the first, providing a sequence of estimates for the fluctuating variable $Q_T$. The
topological susceptibility is estimated from this using
Eqn.(\ref{eqn:topsus}).

One aspect of topological structure that is worth investigating is the size
distribution of the instantons. Instanton size may be calculated from the
peak value of the topological charge density using 
\begin{equation}
q_{\mbox{\scriptsize{peak}}}=\frac{6}{\pi^2\rho^4}.  \label{eqn:itersize}
\end{equation} 
This classical approximation works reasonably
well for large lattice instantons, but for smaller ones whose size is of
the order of the lattice spacing, corrections of $\mathcal{O}(a^2)$ are needed.
The necessary correction factors for $N_c=3$ were calculated by Smith and Teper
by cooling a classical instanton and then parametrising the resulting
relationship between $Q$ and $\rho$~\cite{Smith:1998wt}. The computational
method employed in this study involved reading the peak values of the charge
from the lattice configurations and then applying iterative bisection to find a
value for $\rho$ which satisfied Eqn. (\ref{eqn:itersize}) to within a
predetermined error factor $\epsilon$.

\section{Numerical Results}

Information about the topological structure was extracted using two different
gauge field ensembles.  The first was generated on a $12^3\times24$ lattice with
$\surd\sigma a=0.415(18)$ ($\sigma$ is the string tension)
using $N_f=2$ flavors of Wilson fermion at an inverse
coupling $\beta=1.9$~\cite{Hands:2010gd}. The fermion action included a diquark
source term $aj=0.04$ and the value of the hopping parameter $\kappa=0.168$. The
second ensemble was generated on the same system size, $\beta$ and $j$ using
$N_f=4$, resulting in a significantly finer lattice with $\surd\sigma
a=0.138(4)$ ~\cite{Hands:2011ye}.  This time $\kappa$ was chosen to be 0.158;
both ensembles therefore had a matched pion mass $m_\pi a=0.68(1)$. 
Although we choose to plot several figures in cutoff units $\mu a$, the
horizontal axis could therefore equally be regarded as being calibrated in units of
$\mu/m_\pi$, as noted in \cite{Hands:2011ye}.
In both
cases chemical potential was introduced via the standard Hasenfratz-Karsch
prescription~\cite{Hasenfratz:1983ba}. The minimum cutoff 
for $F\tilde{F}$ was $q_{cut}=0.02a^{-4}$ 
whereas any instantons larger than one third of the spatial extent of the
lattice were ignored.  

\begin{figure}[htb] 
\begin{center}
\epsfig{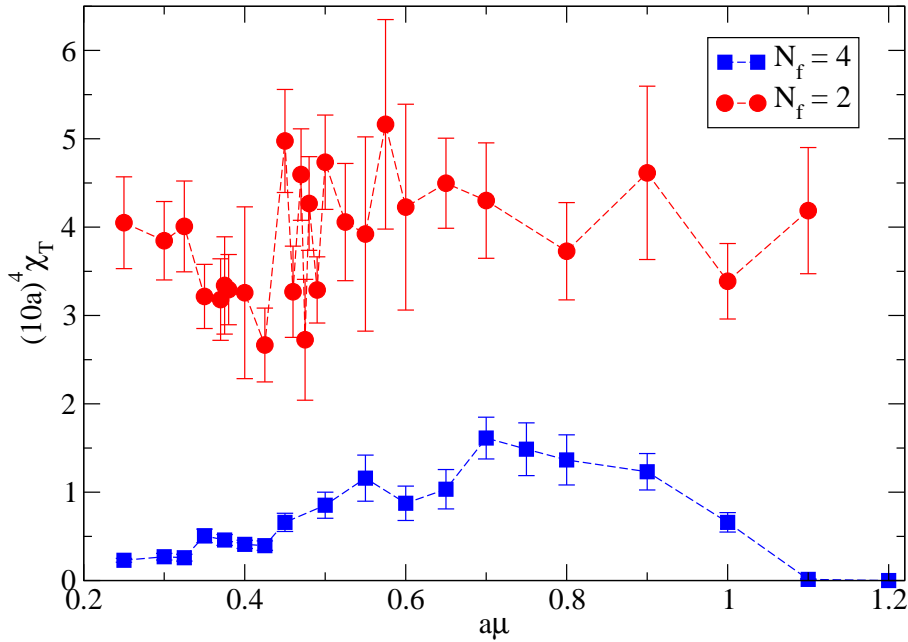}
\end{center}
\vspace{-3mm}
\caption{\em  Topological susceptibility $a^4\chi_T$ versus chemical potential for
$N_f=2$ and $N_f=4$ performing 10 cooling steps.} 
\label{fig:topsus}
\end{figure}
The topological susceptibility of the $N_f=2$ and 4 configurations was measured
across a range $\sim0.25\leq a\mu\leq1.1$. Fig.~\ref{fig:topsus} shows the
behaviour of $\chi_T$ for both theories. In order to verify the validity of our
approach to cooling, the same ensemble was submitted to both 10 and 20 cooling
steps. As the extra cooling had little effect on the signal, we are confident we
are characterising the underlying topology satisfactorily.

In the $N_f=2$ case the signal remains fairly consistent across the range
studied. There are possible signs of some minor downward fluctuations at $a\mu\sim0.4$
and $a\mu\sim0.5$, but beyond $a\mu=0.5$ the data remains flat. The $N_f=4$
data obtained on a finer lattice are more interesting. 
Fig.~\ref{fig:topsus} suggests that
the extra flavors have suppressed the instantons with the peak value of 
$\chi_T^{N_f=4}\sim0.5\chi_T^{N_f=2}$. Moreover, the suppression of
$\chi_T^{N_f=4}$ at the lowest densities is much greater can be explained purely by a change in $N_f$, and its behaviour with increasing
$\mu$ is not as expected if the
relation in Eqn.(\ref{eqn:flavsup}) is correct 
(and assuming $\rho$ to be independent of $N_f$); $\chi(\mu)$
initially increases instead of being suppressed.

\begin{figure}[htb] 
\begin{center}
\epsfig{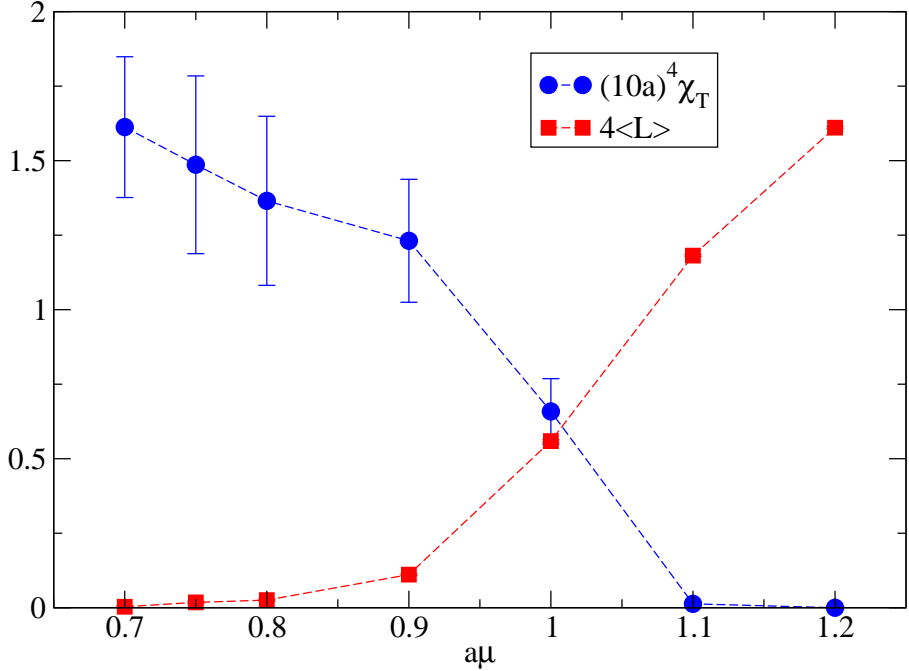}
\end{center}
\vspace{-3mm}
\caption{\em  The suppression of $\chi_T$ coinciding with the rise in
$\left<L\right>$ for $N_f=4$. Note $\langle L\rangle$ has been rescaled for
clarity.} \label{fig:4flcandL} 
\end{figure}
At larger densities the situation changes.
A comparison of $\chi_T^{N_f=4}$ with the 
Polyakov loop $\langle L\rangle$ from~\cite{Hands:2011ye} over the range $0.7<a\mu<1.2$ is shown in
Fig.~\ref{fig:4flcandL}. It illustrates nicely how the fall in $\chi_T$ coincides
with the rise in $\left<L\right>$. 
The Polyakov loop begins to
rise from zero at $a\mu\approx0.8$, whereas $\chi_T$ starts to
fall noticeably just a little later at $a\mu\approx0.9$.
Ref.~\cite{Hands:2011ye} identified a ``deconfining'' value of chemical
potential $a\mu_d\approx0.75$ based on the behaviour of $\langle L\rangle$;
Fig.~\ref{fig:4flcandL} suggests deconfinement in dense matter is accompanied by suppression of
topological fluctuations.

\begin{figure}[htb]
\begin{center}
\epsfig{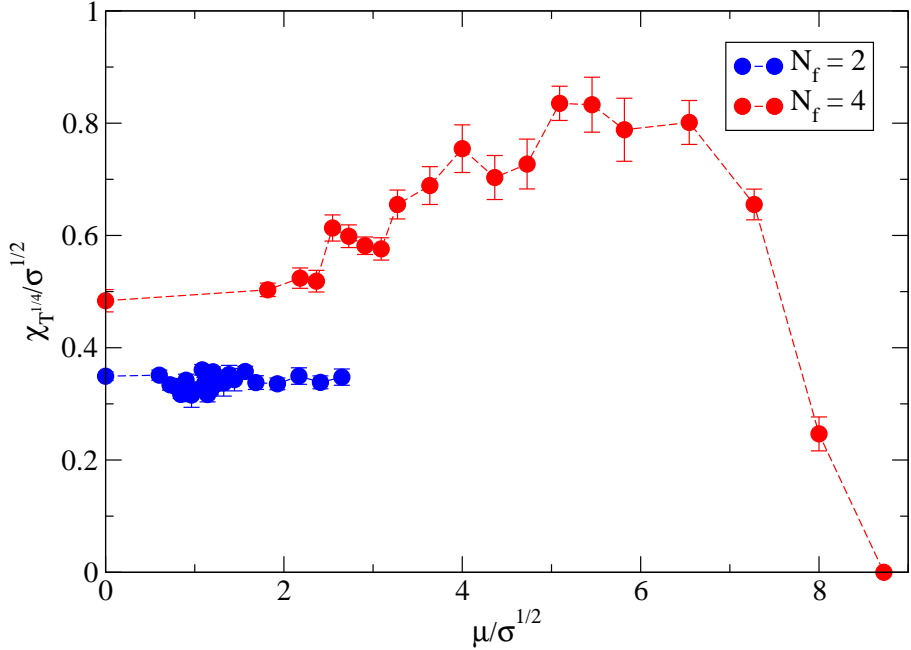}
\end{center}
\vspace{-3mm}
\caption{\em  The dimensionless quantity $\chi_T^\frac{1}{4}/\sigma^\frac{1}{2}$ versus $\mu/\sigma^\frac{1}{2}$  for 2 and 4 flavors.}
\label{fig:dimlesstop}
\end{figure}
It is worth noting that although effort has been made to make a direct
comparison of the $N_f=2$ and $N_f=4$ data sets, in reality they are 
distinct theories. When comparing both
ensembles it is important to remember that the physical volumes (ie. as measured
in string tension units) of the two
lattices differ by a factor of $\sim3^4$. Hence it is difficult to make any useful
quantitative comparison of $\chi_T$ for the two theories directly. It is more
useful to rescale $\chi_T$ in each case as some dimensionless parameter and then
make a comparison of the two.  
Fig.~\ref{fig:dimlesstop} shows the topological susceptibility rescaled and
plotted as the fourth root of $a^4\chi_T$ divided by the square root of the string
tension $a^2\sigma$. From this it is possible to compare the results to
(\ref{eq:WV}),
which implies $\chi_T^{1/4}=180$ MeV. For $N_f=2$,
$\chi_T^{1/4}/\sigma^{1/2}=0.3493\pm0.0076$, and for $N_f=4$,
$\chi_T^{1/4}/\sigma^{1/2}=0.4837\pm0.0198$. Assuming $\sigma=(440\mbox{MeV})^2$
leads to 
\begin{equation} 
\chi_T^\frac{1}{4} = \left\{ \begin{array}{ll} 156\pm3\mbox{MeV}
&  N_f=2;\\ 213\pm9\mbox{MeV} &  N_f=4.\end{array} \right.  
\label{eq:chiMeV}
\end{equation} 
Both
results are in the range suggested by the Witten-Veneziano formula. That the
value for $N_f=2$ is smaller than the $N_f=4$ value by about 25\%
hints that much of the topological information is being missed due to the
coarseness of the lattice.

\begin{figure}[htb]
\begin{center}
\epsfig{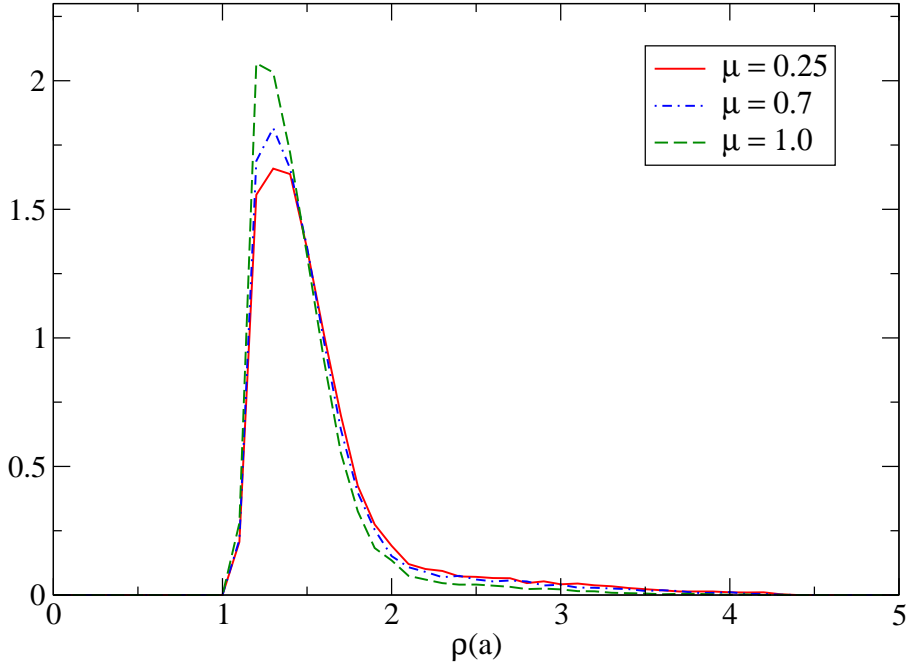}
\end{center}
\vspace{-3mm}
\caption{\em  Normalized instanton size distribution at three different chemical potentials for $N_f=2$. The average size $\bar{\rho}\simeq0.3$ fm.}
\label{fig:2flsizdist}
\end{figure}
\begin{figure}[htb]
\begin{center}
\epsfig{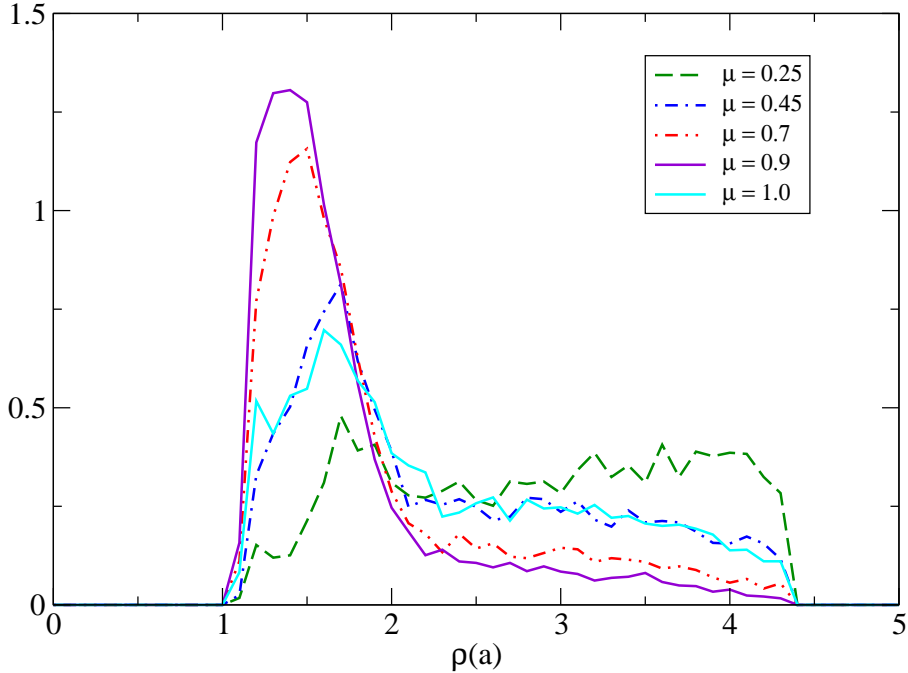}
\end{center}
\vspace{-3mm}
\caption{\em  Behaviour of the normalized $N_f=4$ instanton size distribution at various $\mu$ values.}
\label{fig:4flsizdis}
\end{figure}
The instanton size distribution was plotted for both the two and four flavor
configurations. To compensate for inequalities in the sizes of the data sets, each distribution curve is normalized, keeping the area under the curve consistent. Using the same scale determination that led to
Eqn.~(\ref{eq:chiMeV}), for $N_f=2$ (see Fig.~\ref{fig:2flsizdist}) at $a\mu=0.25$,
the majority lay within the range $0.18\leq\rho\leq0.5$fm, with the average size
being $\bar{\rho}\sim 0.28$fm. This compares well with the phenomenologically
derived value $\bar{\rho}\approx0.3$ fm~\cite{Shuryak:1997vd}. The sharp cutoff
at $\rho=a$ is where instantons smaller than this `fall through' the lattice and
do not contribute. Similar distribution curves are also plotted for higher
values of $\mu$ to see if there is any effect with increasing density. While
there is a hint that larger instantons are slightly suppressed at larger $\mu$,
no significant $\mu$-dependence is observed and all the curves
are qualitatively the same.

By contrast, Fig.~\ref{fig:4flsizdis} shows how the instanton size distribution
evolves with
$\mu$ for $N_f=4$. At $a\mu=0.25$ the distribution is fairly
uniform. As $\mu$ increases the number of large instantons falls as 
that of smaller-sized instantons rises, and the distribution becomes
taller and narrower, with a peak at $a\mu=0.9$ of $\bar{\rho}\sim1.4$fm. For $a\mu=1.0$ the curve has rapidly flattened and has a very similar profile to that for $a\mu=0.45$. The
prevalence of small-sized instantons drives down the average instanton size. The
cutoff at large $\rho$ for $a\mu=0.25,0.45$ results from the constraint on 
the maximum possible instanton size. Such a filter on $\rho$ is needed to minimize
finite volume effects and to prevent instantons overlapping one another. As
$\mu$ increases, $\bar{\rho}$ decreases and this becomes less of an issue but at
low $\mu$, where there is a greater number of large-sized instantons, it seems
likely that some topological information is lost due to the IR cutoff.

\begin{figure}[htb]
\begin{center}
\epsfig{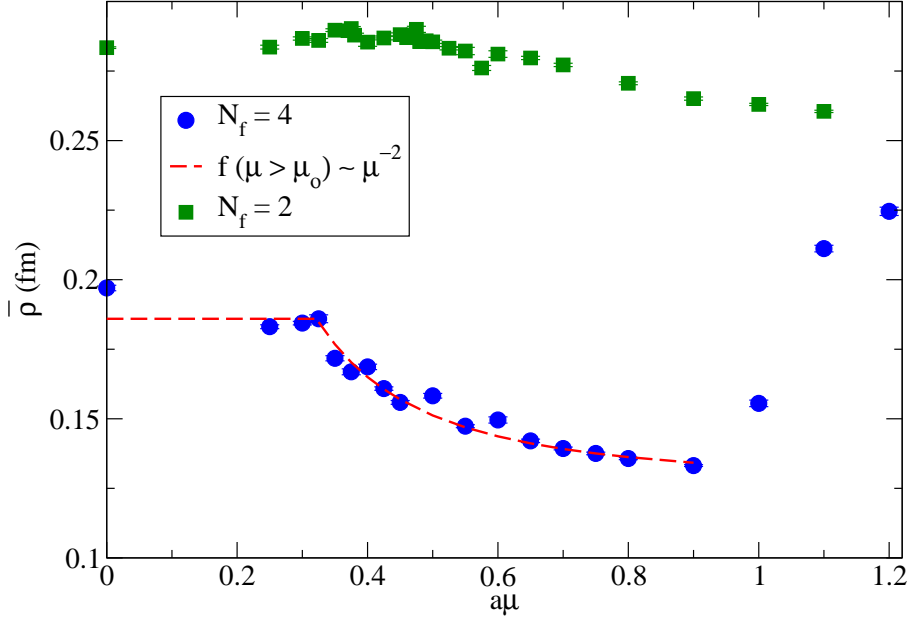}
\end{center}
\vspace{-3mm}
\caption{\em  Average instanton size versus chemical potential for $N_f=2$ and for $N_f=4$ fitted with function $f(\mu)\propto\mu^{-2}$ for $\mu_o<\mu<\mu_d$.}
\label{fig:sizfit}
\end{figure}
\begin{figure}[htb]
\begin{center}
\epsfig{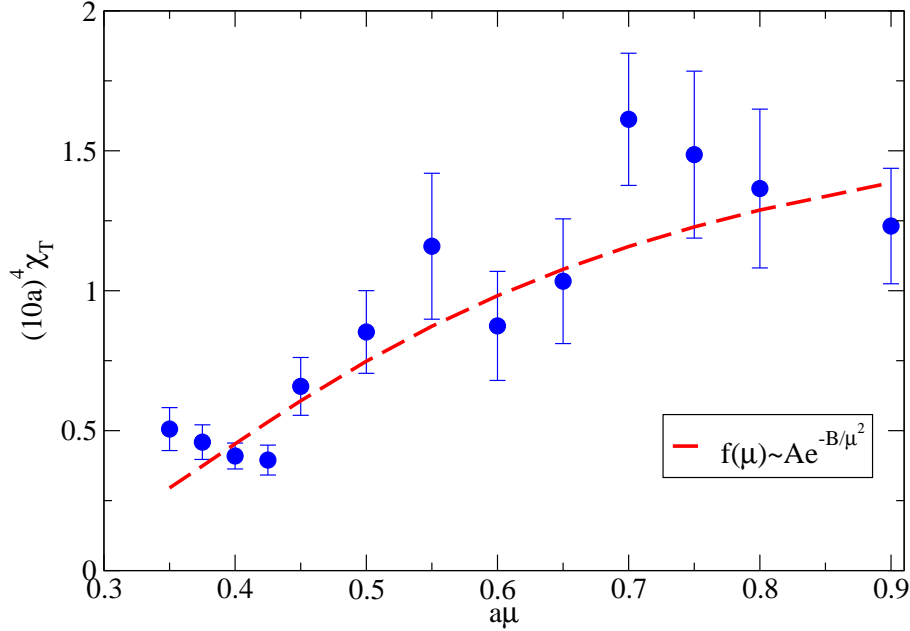}
\end{center}
\vspace{-3mm}
\caption{\em  $\chi_T(\mu)^{N_f=4}$ for $\mu\in(\mu_o,\mu_d)$, 
along with a fit to $f=A\exp\left(-B/\mu^2\right)$.}
\label{fig:chifit}
\end{figure}
When $\bar\rho$ is plotted as a function of $\mu$ (see Fig.~\ref{fig:sizfit}) it is
apparent that the $N_f=4$ instantons are shrinking between onset and
deconfinement, 
after which there is a sharp rise in size. This rise coincides with the rapid
flattening of the $a\mu=1.0$ distribution curve recorded in
Fig.~\ref{fig:4flsizdis}. In the range $\mu_o<\mu<\mu_d$, empirically
$\bar\rho\propto\mu^{-2}$ and this behaviour is plotted along with the data.
This contrasts markedly with the $N_f=2$ data, where $\rho$ appears to be almost
$\mu$-independent. At most there is a very gentle monotonic fall in $\bar{\rho}$
for $\mu>\mu_o$. A consideration of the behaviour of the size distribution in
Fig.~\ref{fig:4flsizdis} suggests why that might be. The most significant
$\mu$-dependent effects are seen for small-sized instantons with
$\rho<3a^{N_f=4}\approx a^{N_f=2}$. Thus, the $N_f=2$ lattice may well be too
coarse for this detail to be seen.

It is also of interest to compare the behaviour of $\rho(\mu)$ for
$N_f=4$ with the predictions of thermal field theory.
The perturbative result (\ref{eqn:flavsup})
of~\cite{Shuryak:1982hk} implies that, for constant
$\rho$ $n$ is suppressed by a factor
$e^{-N_f\rho^2\mu^2}$ as density is increased. Thus, $\chi_T$ should be
suppressed by increasing $\mu$. In Fig.~\ref{fig:topsus}  $\chi_T^{N_f=4}$
rises as $\mu$ increases, which seems to be incompatible with the
perturbative result. However, if we take into account the non-perturbative
information on $\rho(\mu)\propto\mu^{-2}$
in Eqn.(\ref{eqn:flavsup}) we find 
\begin{equation}
n(\mu)\propto\exp(-N_f(\rho(\mu)\mu)^2) =A\exp\left(-\frac{B}{\mu^2}\right).
\label{eqn:risingchi} 
\end{equation} 
The parameter $A$ includes $n(\mu=T=0)$
plus thermal contributions due to the fact that we are working at
low but non-zero $T$. The parameter $B$ includes the constant of proportionality
for instanton size as a function of $\mu$ multiplied by a factor of $N_f$.
When a function of this form is plotted along with $\chi_T^{N_f=4}$ 
in Fig.~\ref{fig:chifit} in the range $\mu_o<\mu<\mu_d$
we see a fair correspondence between the two. The
best fit is found with $A=1.8255$ and $B=0.2231$. 
This suggests that the enhanced topological fluctuations observed in baryonic
matter at moderate density are a direct result of the $\mu$-dependence of the
instanton scale size.

\section{Conclusion}
In this Letter 
we have presented the first exploratory study of topological fluctuations of
non-abelian gauge fields in cold dense baryonic matter, using ensembles
generated for a range of $\mu$ with both $N_f=2$ and $N_f=4$.

For $N_f=2$
the topological susceptibility remained flat across the whole range studied. It is
likely that the lattice is too coarse to be able to capture the topological
detail adequately. The fact that for the two flavor ensemble $\chi_T$ was
measured to be $\chi_T\simeq(150\mbox{ MeV})^4$ suggests that a lot of
topological information is falling through the lattice and being
lost. 

With an increase in the number of flavors to $N_f=4$, the resulting finer lattice 
was able to expose more detail about the distribution
of instanton size and its evolution with $\mu$. Once chemical
potential is increased beyond onset, the instanton size becomes
density-dependent. The large instantons found at low $\mu$
are suppressed as $\mu$ is increased, driving down the
average size to a minimum of $\bar{\rho}\sim0.14$ fm at around $a\mu\sim0.9$. 
The smaller instantons result in reduction in screening of
topological charge fluctuations, so that the topological susceptibility
$\chi_T$ initially rises for $\mu\geq\mu_o$. In the deconfined phase $\mu\geq\mu_d$, 
however, the average instanton size rises sharply, and $\chi_T$ is suppressed.

While it is tempting to ascribe the differences observed between $N_f=2$ and
$N_f=4$ entirely to the different lattice spacings, as measured in string
tension units, we should remain mindful that they are two different theories; in
particular the thermodynamics studies of \cite{Hands:2010gd,Hands:2011ye} reveal
that for $N_f=2$ the regime just above onset is weakly-interacting
and dilute, apparently well-described as a non-relativistic Bose gas of
tightly-bound scalar diquarks. By contrast, matter with $N_f=4$ appears
relativistic and strongly-interacting for all $\mu\geq\mu_o$. 
A systematic study of the $\mu$-dependence of topological fluctuations in 
QC$_2$D, therefore, must await
the generation of gauge ensembles
on a finer lattice.

\section*{Acknowledgments}

This project was enabled with the assistance of IBM Deep Computing.
We are grateful to Biagio Lucini and Jon-Ivar Skullerud for their help and
advice. We have also enjoyed discussing our results with Massimo D'Elia and
Ernst-Michael Ilgenfritz.

\end{document}